\documentclass[prb,superscriptaddress, showpacs,preprint,11pt]{revtex4-1}
\usepackage[utf8]{inputenc}
\usepackage{amsmath}
\usepackage{amssymb}

\usepackage{graphicx}
\usepackage{color}

\begin{document}
\title{Interatomic force constants including the DFT-D dispersion contribution }
\author{Benoit Van Troeye}
\email{benoit.vantroeye@uclouvain.be}
 \affiliation{Institute for Condensed Matter and Nanosciences, European Theoretical Spectroscopy Facility, Universit\'{e} catholique de Louvain, Chemin des étoiles 8, B-1348 Louvain-la-Neuve, Belgium}

\author{Marc Torrent}
\affiliation{CEA, DAM, DIF, F-91297 Arpajon, France}

\author{Xavier Gonze}
\affiliation{Institute for Condensed Matter and Nanosciences, European Theoretical Spectroscopy Facility, Universit\'{e} catholique de Louvain, Chemin des étoiles 8, B-1348 Louvain-la-Neuve, Belgium}

\begin{abstract}
 Grimme's DFT-D dispersion contribution to interatomic forces constants, required for the computation of the phonon band structures in density-functional perturbation theory,
 has been derived analytically. The implementation has then been validated with respect to frozen phonons, and applied on materials where weak cohesive forces play
 a major role i.e. argon, graphite, benzene. We show that these dispersive contributions have to be considered to properly reproduce the experimental vibrational
 properties of these materials, although the lattice parameter change, coming from the ground-state relaxation with the proper functional, induces the most important change with respect to a treatment without dispersion corrections.
 In the current implementation, the contribution of these dispersion corrections to the dynamical matrices (with a number of elements that is proportional to the square of the number of atoms) 
 has only a cubic scaling with the number of atoms. In practice the overload with respect to density-functional calculations is small, making this methodology promising to study vibrational
 properties of large dispersive systems.
\end{abstract}


\maketitle

\section{Introduction}

It is now commonly accepted that the most popular exchange-correlation energy functionals - i.e. LDA, GGA, mGGA or hybrid functionals - fail to properly describe nonlocal dispersion effects in density-functional theory (DFT)   
\cite{Martin2004}. These effects play a major role in layered materials and molecular crystals, leading to the inaccurate description of these materials by the above-mentioned functionals \cite{Klimes2012}. 

To overcome this difficulty, several methods have been developed over the past years.
On the one hand, one finds density-dependent methods or even wavefunction-dependent methods, whose implementation cost might be significant. In this category, several have 
found widespread use, like Tkatchenko-Scheffler van der Waals (TS-vdW) corrections \cite{Tkatchenko2009}, which adds to the DFT result a term  
depending both on the geometry of the system and on the electronic density, or vdW-DF methods \cite{Rydberg2003, Dion2004, Lee2010} which add to the exchange-correlation a nonlocal term to treat the dispersive effects.
 The Silvestrelli approaches (vdW-WF) \cite{Silvestrelli2008,Ambrosetti2012} use maximally
 localized Wannier functions to estimate the vdW correction to KS energy. The random phase approximation \cite{Furche2001,Fuchs2002,Ren2012} has also shown encouraging
 results to take into account of these interactions, although it still suffers from a large computational time overhead \cite{Lu2009}.

On the other hand, simpler density-independent dispersion corrections, that include S. Grimme's DFT-D methods, have been developed, with a quite significant impact as well \cite{Grimme2006, Grimme2010, Grimme2010b}. 
In this case, the correction only depends on the nuclei positions and on the approximation for the exchange-correlation functional in use. 

In DFT-D2 \cite{Grimme2006}, a simple pair-wise term is added to the DFT energy. It exhibits a long-range behavior $C_{6,ij}/R_{ij}^{6}$ where $C_{6,ij}$ is the 
dispersion coefficient and $R_{ij}$ is the distance between the atoms.
Although yielding better agreement with the experiments than DFT for most non-covalently bound systems, this method only considers one coefficient for each chemical pair and thus may not be able to catch
fundamental trends of these interactions. For example, the dispersion coefficient can vary by a factor of two in the case of armchair carbon nanotubes depending of their size \cite{Gobre2013}. 

In the more sophisticated DFT-D3 \cite{Grimme2010}, the problem is tackled with the use of environment-dependent dispersion coefficients. These coefficients are interpolated on tabulated supporting points which
have been computed beforehand in TDDFT. A 3-body term -also referred as the Axilrod–Teller–Muto nonadditive term \cite{Axilrod1943,Muto1943}- can also be taken into account.
Finally, DFT-D3 with Becke-Johnson damping -DFT-D3(BJ) \cite{Grimme2010b}- is a variant of the previously 
introduced DFT-D3 method that uses another expression for the damping, which removes the undesired short-range divergent behavior of the correction.

In all these cases, these corrections have to be taken into account in the computation of atomic derivative-related quantities like forces, 
stresses, interatomic force constants (IFCs), dynamical matrices, or elastic tensor. At variance with vdW-DF or TS-vdW methods, the DFT-D ones do not lead to direct modifications of properties related to the electric field derivatives i.e. Born effective charges, dielectric tensor
or electron-phonon coupling matrix elements, as these corrections are independent of the density, except for the indirect dependence through modification of atomic positions. This is both an advantage and a drawback of these methods, 
as on the one hand the complexity of the equations and their implementation
remains quite low but on the other hand, some effects could be lost by neglecting the density dependence of the vdW corrections. 

Although it is possible to compute all the previously introduced global quantities in the case of DFT with finite difference
techniques, the computations can become quite demanding, especially for dynamical matrices at small wavevectors, which require the use of large supercells.
A more elegant way to calculate these quantities is achieved in density-functional perturbation theory (DFPT) \cite{Baroni1991,Gonze1997,Gonze1997b,Abinit2005}, which benefits from lower computation time and more easily achieved precision. 
This formalism has also be extended to strain perturbations \cite{Hamann2005, Hamann2005b} for the computation of relaxed elastic, piezoelectric and internal strain coupling parameters.

In this article, we will show how the DFT-D pair-wise contribution to the Fourier transform of the IFCs at any wavevector of the reciprocal space,
required for the computation of the phonons frequencies and eigendisplacements, can be derived in a similar scheme. We will neglect the effect of the 3-body term,
as discussed later. We are aware of the existence of a DFPT phonon implementation
with the vdW-DF methodology \cite{Sabatini2012}, although it has not been released to our knowledge. The DFT-D2 contributions to the IFCs have also 
been reported in the \textsc{Gulp} software \cite{Gale1997,Fang2014}. 
Except for these implementations and the related publications, the litterature reports some frozen-phonon computations with 
dispersion corrections in periodic systems \cite{Otero2013, Reilly2013,Reilly2014,Fedorov2015,Ling2015}.
Our method will be afterwards applied to three materials incorrectly described by DFT in GGA i.e. 
argon, graphite and benzene. 

It has to be reminded that phonon modes are quite sensitive to the geometry of the system. As these DFT-D corrections play an important role for the ground-state geometry, two effects on the phonons
have to be distinguish when the vdW corrections are taken into account: an indirect effect, related to the change of geometry of the ground state, and a direct effect which is related to
the contribution of these corrections to the IFCs.
It will be shown that although the largest effect of DFT-D methods on the phonon frequencies originates from geometrical modifications, the dispersion corrections to the IFCs cannot be neglected.

In \textsc{Sec. \ref{Section1}}, the theoretical background will be presented, with some detailed derivations in the supplementary informations.
The DFT-D contribution
to the dynamical matrix can be implemented with an $O(N_{at}^3)$ scaling. Taking into account the prefactor of the calculation, and the normal cost and scaling of usual DFPT calculations of a dynamical matrix,
the associated computational overhead is negligible, whatever the size of the system. 
We will also discuss briefly in \textsc{Sec. \ref{Implementation2}} the implementation and show that an excellent agreement can be obtained between our implementation and frozen-phonon calculations.
Finally, in \textsc{Sec. \ref{Section2}},
we will present the results obtained with our implementation on specific materials.

\section{Theory} \label{Section1}

In all DFT-D methods, a density-independent pair-wise contribution $E_{disp}^{(2)}$ is added to the DFT energy to treat the dispersion. 
In the case of periodic systems, this contribution to the energy of cell $\boldsymbol{\tau}$ can be expressed as
\begin{equation}
 E^{(2)}_{disp}(\boldsymbol{\tau}) = - \sum_{i}^{N_{at}} \sum_{\boldsymbol{\tau'}} \sum_{j}^{N_{at}} C^{\boldsymbol{\tau}\boldsymbol{\tau'}}_{6,ij}(\left\{\boldsymbol{R}\right\})
 f(R^{\boldsymbol{\tau}\boldsymbol{\tau'}}_{ij}), \label{Equation1}
\end{equation}
where $\boldsymbol{\tau'}$ is a cell replica index, $i$ and $j$ are indices of atoms in the primitive cell,
$C_{6,ij}^{\boldsymbol{\tau}\boldsymbol{\tau'}}(\left\{\boldsymbol{R}\right\})$ is the dispersion coefficient between $i$ in cell $\boldsymbol{\tau}$ and $j$ in cell $\boldsymbol{\tau'}$, function in DFT-D3 and DFT-D3(BJ)
of the whole set of atomic positions $\left\{\boldsymbol{R}\right\}$,
and $f$ is the analytical function used to describe the dispersion which depends on the DFT-D method used, on the chemical species of $i$ and $j$, as well as of the distance between the two considered atoms $R^{\boldsymbol{\tau}\boldsymbol{\tau'}}_{ij}$.
For example, in DFT-D3 it takes the form
\begin{equation}
 f^{\text{D3}}(R^{\boldsymbol{\tau}\boldsymbol{\tau'}}_{ij}) = \frac{1}{2} \left[ s_6 \frac{f_{dmp,6}(R_{ij}^{\boldsymbol{\tau}\boldsymbol{\tau'}})}{\left(R_{ij}^{\boldsymbol{\tau}\boldsymbol{\tau'}}\right)^6} + \right.
  \left. 3 s_8 \sqrt{Q_i}\sqrt{Q_j} \frac{f_{dmp,8}(R_{ij}^{\boldsymbol{\tau}\boldsymbol{\tau'}})}{\left(R_{ij}^{\boldsymbol{\tau}\boldsymbol{\tau'}}\right)^8} \right],
\end{equation}
where $s_n$ are coefficients which depend of the exchange-correlation functional used, $f_{dmp,n}$ are the n$^{\text{th}}$-order damping functions used to remove the short-range divergent behavior of the
function, and $Q_i$ are tabulated values expressing the link between lower and higher dispersion coefficients.

As discussed previously, in DFT-D3 and DFT-D3(BJ), the dispersion coefficients depend on the chemical environment around each atom. For sake of brevity, the dependence of this function w.r. to the atomic positions $\left\{\boldsymbol{R
}\right\}$ will be implied for the remaining of this paper. The dispersion coefficients are interpolated between supporting points as follows:
\begin{equation}
 C_{6,ij}^{\boldsymbol{\tau}\boldsymbol{\tau'}} =  \frac{1}{L_{ij}^{\text{tot},\boldsymbol{\tau}\boldsymbol{\tau'}}} \sum_{r_i}^{r_{i,\text{max}}} \sum_{r_j}^{r_{j,\text{max}}} 
 C^{\text{ref}}_{6,ij,r_i r_j} L_{ij,r_i r_j}^{\boldsymbol{\tau}\boldsymbol{\tau'}},
\end{equation}
with $r_i$ the reference for the chemical species of atom i, $r_{i,\text{max}}$ the number of tabulated value available for the considered chemical species (e.g. five in the case of carbon),
\begin{equation}
 L_{ij}^{\text{tot},\boldsymbol{\tau}\boldsymbol{\tau'}} = \sum_{r_i}^{r_\text{max}} \sum_{r_\text{j}}^{r_\text{max}} L_{ij,r_i r_j}^{\boldsymbol{\tau}\boldsymbol{\tau'}},
\end{equation}
and 
\begin{equation}
 L_{ij,r_i r_j}^{\boldsymbol{\tau}\boldsymbol{\tau'}} = e^{-k_3\left[(CN_i^{\boldsymbol{\tau}}-CN^{\text{ref}}_{i,r_i})^2+(CN_j^{\boldsymbol{\tau'}}-CN^{\text{ref}}_{j,r_j})^2\right]}.
\end{equation}

The $CN^{\text{ref}}_{i,r_i}$ and $CN^{\text{ref}}_{j,r_j}$ tensors contain the supporting points for the interpolation, while $C^{\text{ref}}_{6,ij,r_ir_j}$ 
contains the reference values for the dispersion coefficients, which have been computed beforehand in TDDFT \cite{Grimme2010}. 
Finally, $k_3=4$ and $CN_i^{\boldsymbol{\tau}}$ is the coordination number of atom $i$ in cell $\boldsymbol{\tau}$. 
In periodic systems, the coordination number as proposed in the original Grimme's paper was a diverging quantity, as pointed out by Reckien \cite{Reckien2012}. 
The latter author refined the expression as follows :
\begin{equation}
 CN_i^{\boldsymbol{\tau}} = \sum_{j}^{N_{at}} \sum_{\boldsymbol{\tau''}}  \left[1+e^{-k_1\left(k_2 \frac{R_{\text{cov},i}+R_{\text{cov},j}}{R_{ij}^{\boldsymbol{\tau}\boldsymbol{\tau''}}}-1 \right)}\right]^{-1}
  \times f_{dmp,CN}(R^{\boldsymbol{\tau\tau''}}_{ij}),
\label{cn}
\end{equation}\\

with $R_{\text{cov},i}$ being the covalence radius of species $i$, $k_1=16$, $k_2=4/3$ and
\begin{equation}
f_{dmp,CN}(R^{\boldsymbol{\tau\tau''}}_{ij}) = \frac{1}{2} \text{erfc} \left[R^{\boldsymbol{\tau\tau''}}_{ij}-15 k_2 (R_{\text{cov},i}+R_{\text{cov},j})\right].
\end{equation} We use the same expression for our implementation. Note that by translational invariance, all the previous introduced quantities are periodic
and thus can be computed taking $\boldsymbol{\tau}=\boldsymbol{0}$.

In DFT-D3, a three-body correction $E^{(3)}_{disp}(\boldsymbol{\tau})$ is also taken into account for the dispersion; it is computed by summing the partial contribution
of all the triplets of atoms

\begin{widetext}
\begin{equation}
   E^{(3)}_{disp}(\boldsymbol{\tau}) = -\sum_{i}^{N_{at}} \sum_{\boldsymbol{\tau'}} \sum_{j}^{N_{at}} \sum_{\boldsymbol{\tau''}} \sum_{k}^{N_{at}}
   \frac{s_9}{6} C_{9,ijk}^{\boldsymbol{\tau}\boldsymbol{\tau'}\boldsymbol{\tau''}} \frac{1+3 \cos(\alpha) \cos(\beta) \cos(\gamma)}{
   \left(R_{ij}^{\boldsymbol{\tau}\boldsymbol{\tau'}}R_{jk}^{\boldsymbol{\tau'}\boldsymbol{\tau''}}R_{ki}^{\boldsymbol{\tau''}\boldsymbol{\tau}}\right)^3} 
   f_{dmp,9}\left(\bar{R}_{ijk}^{\boldsymbol{\tau}\boldsymbol{\tau'}\boldsymbol{\tau''}}\right), 
\end{equation}
\end{widetext}
where
\begin{equation}
 C_{9,ijk}^{\boldsymbol{\tau}\boldsymbol{\tau'}\boldsymbol{\tau''}} = - \sqrt{C_{6,ij}^{\boldsymbol{\tau}\boldsymbol{\tau'}}
 C_{6,jk}^{\boldsymbol{\tau'}\boldsymbol{\tau''}}C_{6,ki}^{\boldsymbol{\tau''}\boldsymbol{\tau}}},
\end{equation}
$\alpha$, $\beta$, $\gamma$ are the angles of the triangle formed by the triplet of atoms and $f_{dmp,9}$ is the associated
damping function with $\bar{R}_{ijk}^{\boldsymbol{\tau}\boldsymbol{\tau'}\boldsymbol{\tau''}}$ the geometrical mean distance between the three atoms of the triplet.

In this work, we will neglect this three-body term (with one exception described later). Indeed, despite arising naturally from the theory of Van der Waals interaction,
the use of this 3-body term in practical calculations is still debatable, as it tends for example to worsen the cohesive energy for GGA-PBE compared to pair-wise corrections alone while improving it for HSE06\cite{Moellemann2014}.
In any case, it yields much smaller contributions to the energy and its derivative than the pair-wise term (about 5\% of the total dispersion contribution to
the binding energy and neglegible role on geometry optimization \cite{Moellemann2014}). 
We nevertheless performed finite differences on this three-body term in order to estimate its contribution to the IFCs. The results are presented in \textsc{Sec. \ref{Section2}}.

For the first-order perturbation (forces and stresses), it is shown in \textsc{S.I.} that
the pair-wise contribution to forces and stresses scale as $O(N_{at}^2)$ while the one of the three-body term scales as $O(N_{at}^3)$ \footnote{It is already the case
in Grimme's software (www.thch.uni-bonn.de/tc/index.php).}. 

The pair-wise dispersion contribution to the IFCs is given by
\begin{equation}
 C^{disp}_{\kappa\alpha,\kappa'\beta}(\boldsymbol{0},\boldsymbol{b}) = \frac{\partial^2 E_{disp}}{\partial R_{\kappa \alpha}^{\boldsymbol{0}}\partial R_{\kappa' \beta}^{\boldsymbol{b}}}.
\end{equation}
where $\alpha, \beta$ corresponds to the directions along which the atoms are moved. 

\begin{figure}[h]
\hspace{-0.4cm}\includegraphics[width=0.5\textwidth]{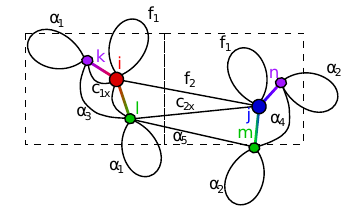}
\caption{\label{Schema} Schematic representation of all the DFT-D contribution to the IFCs detailed in \textsc{Eq. \ref{full}} and in \textsc{S.I.}.
There are nine distinct terms labeled $f_1$, $f_2$, $\alpha_1$ to $\alpha_5$, $c_{1x}$ and $c_{2x}$;
i and j refer to the atom considered in the pairwise term. k and l are the atoms
that contribute to the coordination number of atom i, while m and n contribute to the coordination of atom j. A plain black line that connects two atoms refers to the
second derivative of i-j pair contribution to the dispersion energy with respect to the displacements of these two atoms.}
\end{figure}

From \textsc{Eq. \ref{Equation1}}, one can see that the dispersion contribution to IFCs for the pair-wise term will be linked to the second derivative of $f(R_{ij}^{\boldsymbol{\tau}\boldsymbol{\tau'}})$, and in the case of DFT-D3 and DFT-D3(BJ), 
to the second derivative of $C^{\boldsymbol{\tau}\boldsymbol{\tau'}}_{6,ij}$ and to the cross derivatives of $f(R_{ij}^{\boldsymbol{\tau}\boldsymbol{\tau'}})$ and $C^{\boldsymbol{\tau}\boldsymbol{\tau'}}_{6,ij}$ with respect to two atomic displacements. 
The derivatives of $C^{\boldsymbol{\tau}\boldsymbol{\tau'}}_{6,ij}$ are themselves related to the derivatives of $CN^{\boldsymbol{\tau}}_i$ and $CN^{\boldsymbol{\tau'}}_j$. 
These dispersion contributions to the IFCs include thus many terms: all of them, as well as which atoms are involved, are shown schematically in \textsc{Fig. \ref{Schema}}.

Mathematically, the full
contribution is given by

\begin{multline}
 C^{disp}_{\kappa\alpha,\kappa'\beta}(\boldsymbol{0},\boldsymbol{b}) = 
 \underbrace{C^{disp,f1}_{\kappa\alpha,\kappa'\beta}(\boldsymbol{0},\boldsymbol{b}) +
 C^{disp,f2}_{\kappa\alpha,\kappa'\beta}(\boldsymbol{0},\boldsymbol{b})}_{f \text{ derivatives}} + 
 \underbrace{C^{disp,c1\times}_{\kappa\alpha,\kappa'\beta}(\boldsymbol{0},\boldsymbol{b}) +
 C^{disp,c2\times}_{\kappa\alpha,\kappa'\beta}(\boldsymbol{0},\boldsymbol{b})}_{f\times \text{CN}_i \, \& \, f \times\text{CN}_j \text{ derivatives}}+\\
 \underbrace{C^{disp,\alpha1}_{\kappa\alpha,\kappa'\beta}(\boldsymbol{0},\boldsymbol{b}) +
 C^{disp,\alpha2}_{\kappa\alpha,\kappa'\beta}(\boldsymbol{0},\boldsymbol{b}) +
 C^{disp,\alpha3}_{\kappa\alpha,\kappa'\beta}(\boldsymbol{0},\boldsymbol{b}) +
 C^{disp,\alpha4}_{\kappa\alpha,\kappa'\beta}(\boldsymbol{0},\boldsymbol{b}) +
 C^{disp,\alpha5}_{\kappa\alpha,\kappa'\beta}(\boldsymbol{0},\boldsymbol{b})}_{\text{CN}_i \, \& \, \text{CN}_j \text{ derivatives}}. \label{full} \\
\end{multline}

The discrete Fourier transform of this last expression for a $\boldsymbol{q}$-vector of the reciprocal space,
\begin{equation}
 \overset{\sim}{C} ^{disp}_{\kappa\alpha,\kappa'\beta}(\boldsymbol{q}) = \sum_{\boldsymbol{b}} C^{disp}_{\kappa\alpha,\kappa'\beta}(\boldsymbol{0},\boldsymbol{b}) e^{i \boldsymbol{q}.\boldsymbol{R_b}},
\end{equation}
can be added to the dynamical matrix calculated in DFPT for the computation of the phonon frequencies and eigenmodes
of the crystal under study. 

The full theoretical derivation of this last Fourier transform can be found in \textsc{S.I.}. It will be shown that,
for the pair-wise term, the DFT-D contribution to the dynamical matrix scales only as $O(N_{at}^3)$. 

To validate these expressions, frozen-phonon computations were realized and are presented in \textsc{Sec. \ref{Implementation2}}. This theoretical framework was then applied on specific materials
to compute their phonon frequencies with inclusion of DFT-D contributions; the results are presented and discussed in \textsc{Sec. \ref{Section2}}. 

\section{Implementation and tests} \label{Implementation2}

DFT-D methods have been implemented within the \textsc{Abinit} software \cite{Abinit2005,Abinit2009} for both ground state and for atomic response functions. 
As the previously-introduced contributions to the energy, forces, stresses and interatomic force constants can not be computed for an infinite number of cell replica, 
a tolerance is used to define the number of cells to be considered in the DFT-D correction. For the computation of the coordination number, the cut-off radius was set to 106 \AA $\,$
while for the pair-wise term, a tolerance on the energy of $10^{-12}$ Ha is used \footnote{Only the pairs for which the contribution to DFT-D energy is higher than the tolerance are considered.}.

First, our implementation has been validated with respect to Grimme's code\footnote{www.thch.uni-bonn.de/tc/index.php}. 
We tested graphite (AB stacking) with the GGA-PBE functional and with the in-plane and out-plane lattice parameters used as the relaxed DFT-D3 ones, i.e. 2.46 \AA $\,$ and 6.96 \AA.
The computation with \textsc{Abinit} of the pair-wise and 3-body dispersion energies gave, respectively, -13.40 mHa and 1.286 mHa for DFT-D3. These values
have to be compared to -13.42 mHa and 1.284 mHa obtained with Grimme's code. For DFT-D2 and DFT-D3(BJ), we obtain for the pair-wise term -16.826 mHa and -22.00 mHa in \textsc{Abinit}, respectively, while we get
-16.824 mHa and -22.03 mHa with Grimme's code. The remaining discrepancies for DFT-D3 and DFT-D3(BJ) can be explained by the absence of cut-off for the coordination number in Grimme's code while present in our implementation. 
These tests validate the implementation of these DFT-D methods inside the \textsc{Abinit} software.

Second, in order to validate the dispersion contribution to the IFCs in reciprocal space, we computed this quantity with DFPT and
frozen phonons at specific q-points using supercells. We used relative atomic displacements of $10^{-7}$ and a first-order finite 
difference technique on the forces to get these values.  

The comparison for graphite of the DFT-D3 contribution to the IFCs with our implementation and 
with finite difference is illustrated in \textsc{Tab. \ref{TabComp}}, with the 3-body term being neglected in the two cases.
As one can see, agreement up to 6 digits can be easily achieved. 
This confirms the validity of the previously-introduced mathematical derivations.
\begin{table}[h]
\begin{center}
\begin{tabular}{l c c}
\hline \hline
& Frozen phonons & DFPT \\
& DFT-D IFCs [mHa] & DFT-D IFCs [mHa] \\
\hline \\[-0.1cm]
$\mathbb{R}[\tilde{C}_{1313}^{disp}(\boldsymbol{\Gamma})]$ & $\mathbf{-10.9750020}73$ & $\mathbf{-10.9750020}80$ \\                                              
$\mathbb{R}[\tilde{C}_{1313}^{disp}(\boldsymbol{A})]$ &      $\mathbf{-16.681563}853$ & $\mathbf{-16.681563}203$ \\
$\mathbb{R}[\tilde{C}_{1313}^{disp}(0,0,1/3)]$ &  $\mathbf{-15.3102516}33$ & $\mathbf{-15.3102516}46$ \\ 
$\mathbb{R}[\tilde{C}_{1323}^{disp}(0,0,1/3)]$ &   $\mathbf{13.009670}591$ &  $\mathbf{13.009670}637$ \\
$\mathbb{I}[\tilde{C}_{1323}^{disp}(0,0,1/3)]$ &  0.0000000000 &  0.0000000000 \\
 \\[-0.1cm]
\hline \hline
\end{tabular}
\end{center}
\caption{Validation of our implementation by comparison of dispersion contribution to IFCs (reduced coordinates) in DFT-D3
computed by frozen phonons and by DFPT. The q-vector is given in reduced coordinates. \label{TabComp}}
\end{table}


Finally, we examined the influence on the IFCs of the 3-body term, if included, compared to the one of the pair-wise term thanks to the same finite difference technique. 
The energy tolerance was set for this 3-body term to $10^{-11}$ Ha and computations were performed at the DFT-D3 (pair-wise only) geometry of graphite. 
We obtained for the specific three-body contribution $\tilde{C}^{\text{3-bt}}_{1313}(\boldsymbol{\Gamma})\approx$ -1.6\,mHa, around 15\% of the pair-wise contribution to the IFCs (equal to -10.975\,mHa in our case). 
In consequence, neglecting this 3-body contribution to the IFCs is an approximation with an impact similar to neglecting it at the total energy or geometry relaxation levels (that we do anyway in this article).

\section{Applications} \label{Section2}

In this section, we present some results obtained with our implementation. We took three well-known materials that require dispersion corrections to be properly described with an ab initio method: 
solid argon, graphite and benzene. All computations are performed in DFT/DFPT and with the software \textsc{Abinit}. 

The GGA-PBE approximation \cite{Perdew1996} was adopted for the exchange-correlation functional in addition of the DFT-D methods. As already mentioned, we neglect the 3-body contribution for ground-state and vibrational properties.
An energy cut-off smearing  \cite{Bernasconi1995} of 0.5 Ha is used and geometry optimizations were carried on until the forces on each
atom were smaller than 10$^{-8}$ Ha/Bohr. Phonon frequencies were computed at relaxed lattice parameters. We use such strict relaxation criterion due to the weak nature of the dispersive forces. 
Concerning the convergence criteria with respect to the plane-wave cut-off energy and to the density of the Monkhorst grid \cite{Monkhorst1976}, we required a precision better than 0.2\% on the lattice parameters and
of $1$ cm$^{-1}$ on the low-frequency modes, referred as ``lattice'' modes in this paper. Further computational details, like convergence parameters for each material under study, are given in the \textsc{S.I.}.

We use the following definition for the cohesive energy by unit cell of the crystal

\begin{equation}
 E_{coh} = \frac{E_{solid}}{N} - E_{gas},
\end{equation}

where $E_{solid}$ is the total energy computed at relaxed position, $E_{gas}$ the total energy computed when the atoms, layers or molecules are at least 16 \AA $\,$ away to their closest neighbor and N the number of
molecules by primitive cell in the crystal.

It is finally important to mention that we neglect the effect of the zero-point motion on the cohesive energy, equilibrium lattice parameters and phonon frequencies 
in our computation, although anharmonic effects may play an important role in molecular crystals. 

\subsection{Argon}
The isotope 36 of argon crystallizes in the FCC spatial arrangement at around 84K under normal conditions. Due to the dispersive nature of the long-range interactions between the Ar atoms, DFT-PBE fails to describe properly this system: it predicts a lattice constant of 5.95 \AA,
quite off compared to the experimental value of 5.30017 \AA \cite{Peterson1966} at 4.25K. DFT-D2,-D3 and -D3(BJ) give 5.37 \AA, 5.56 \AA, 5.48  \AA 
$\,$ for the lattice parameter, respectively, in better agreement with the experiments. 
Our DFT-D3(BJ) implementation gives a cohesive energy of -88 meV, consistent with the value of -87 meV reported in the literature with the same method in \textsc{VASP} \cite{Moellemann2014}.

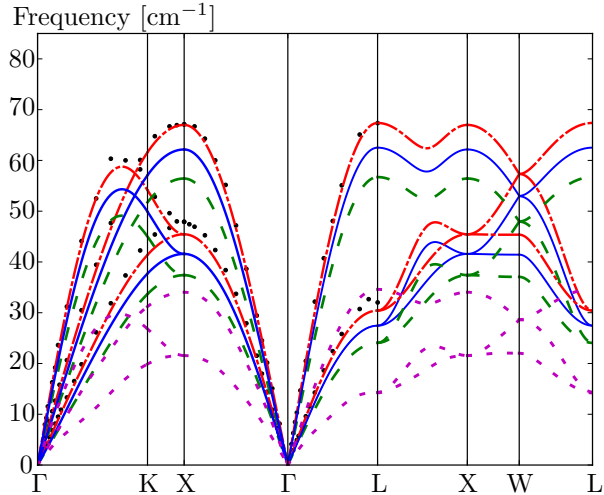
\begin{figure}[h]
\hspace{-0.3cm}\begingroup%
  \makeatletter%
  \providecommand\color[2][]{%
    \errmessage{(Inkscape) Color is used for the text in Inkscape, but the package 'color.sty' is not loaded}%
    \renewcommand\color[2][]{}%
  }%
  \providecommand\transparent[1]{%
    \errmessage{(Inkscape) Transparency is used (non-zero) for the text in Inkscape, but the package 'transparent.sty' is not loaded}%
    \renewcommand\transparent[1]{}%
  }%
  \providecommand\rotatebox[2]{#2}%
  \ifx\svgwidth\undefined%
    \setlength{\unitlength}{250bp}%
    \ifx\svgscale\undefined%
      \relax%
    \else%
      \setlength{\unitlength}{\unitlength * \real{\svgscale}}%
    \fi%
  \else%
    \setlength{\unitlength}{\svgwidth}%
  \fi%
  \global\let\svgwidth\undefined%
  \global\let\svgscale\undefined%
  \makeatother%
  \begin{picture}(1,0.75384615)%
    \put(0,0){\includegraphics[width=\unitlength]{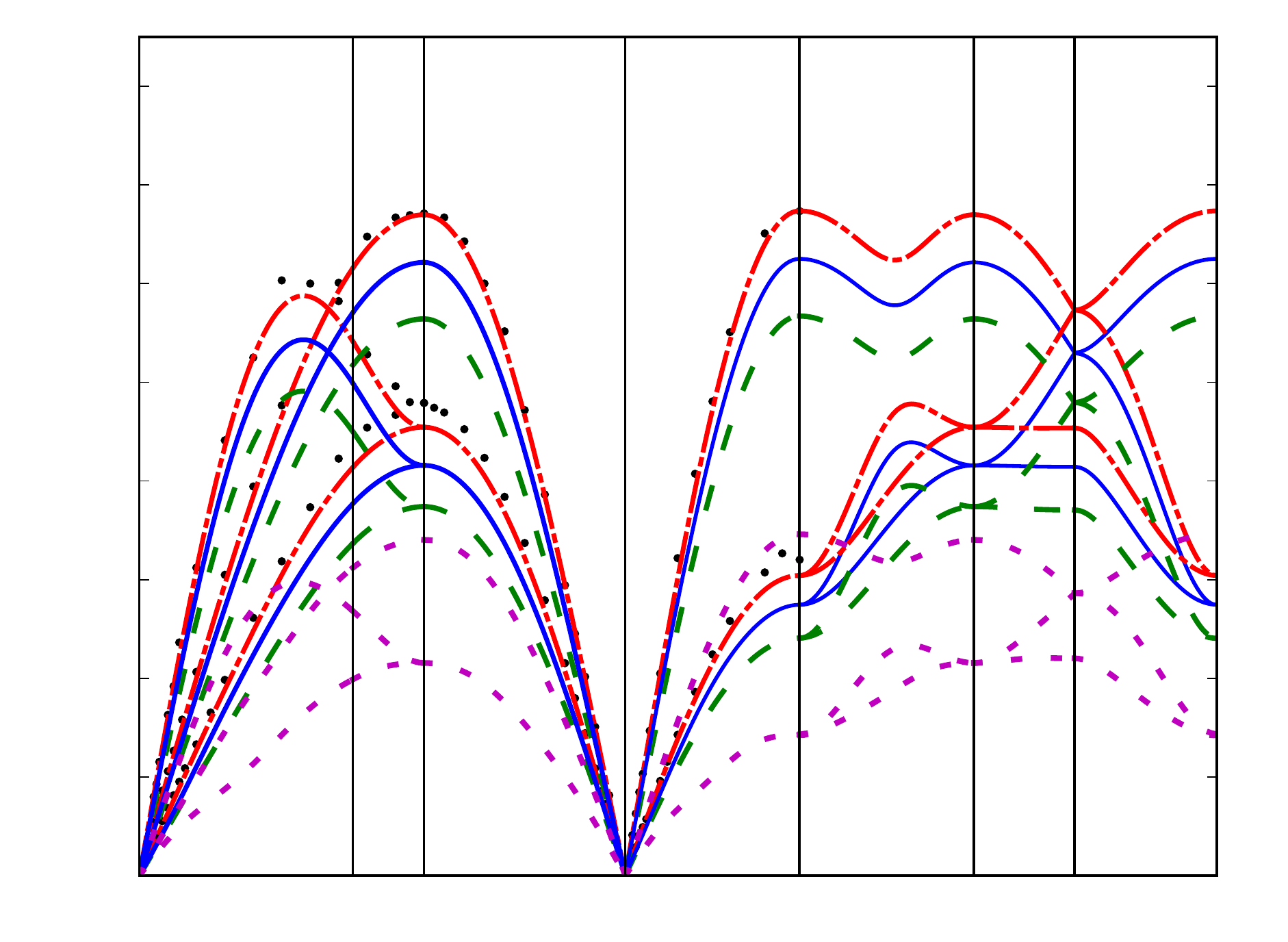}}%
    \put(0.09901709,0.010){\makebox(0,0)[lb]{$\Gamma$}}%
    \put(0.26047863,0.010){\makebox(0,0)[lb]{K}}%
    \put(0.31721368,0.010){\makebox(0,0)[lb]{X}}%
    \put(0.47453504,0.010){\makebox(0,0)[lb]{$\Gamma$}}%
    \put(0.61102222,0.010){\makebox(0,0)[lb]{L}}%
    \put(0.74259658,0.010){\makebox(0,0)[lb]{X}}%
    \put(0.81502222,0.010){\makebox(0,0)[lb]{W}}%
    \put(0.93619316,0.010){\makebox(0,0)[lb]{L}}%
    \put(0.0818051,0.03877402){\makebox(0,0)[lb]{0}}%
    \put(0.0651650103,0.11772427){\makebox(0,0)[lb]{10}}%
    \put(0.06519575316,0.19567521){\makebox(0,0)[lb]{20}}%
    \put(0.06519583333,0.27162393){\makebox(0,0)[lb]{30}}%
    \put(0.06519527248,0.34757436){\makebox(0,0)[lb]{40}}%
    \put(0.06519583333,0.42352479){\makebox(0,0)[lb]{50}}%
    \put(0.06519569983,0.49447521){\makebox(0,0)[lb]{60}}%
    \put(0.06519594017,0.579542564){\makebox(0,0)[lb]{70}}%
    \put(0.06519564632,0.65237607){\makebox(0,0)[lb]{80}}%
    \put(0.07108769,0.705547009){\makebox(0,0)[lb]{Frequency [cm$^{-1}$]}}%
  \end{picture}%
\endgroup%
\caption{\label{PBS_Argon} Phonon band structure of solid $^{36}$Ar computed with the different DFT-D corrections at the corresponding relaxed lattice parameter. \textsc{Dotted purple:} DFT-PBE without DFT-D correction; \textsc{Dashed green:} DFT-D3; \textsc{Solid blue:} 
\textsc{DFT-D3(BJ)}, \textsc{Dash-dot red:} \textsc{DFT-D2}. In each case, the computations were performed at the corresponding relaxed lattice parameters. 
Experimental data from \textsc{Ref.} \onlinecite{Fujii1974} are also shown (black dots). }
\end{figure}

The phonon band structures of Ar computed with the different DFT-D methods, are presented in \textsc{Fig. \ref{PBS_Argon}}. In each case, the computations were performed at the corresponding relaxed lattice parameters
and imposing the mass of the argon as 36 amu. Experimental measurements performed at 10 K by Fujii and coworkers \cite{Fujii1974} are also shown.

At first sight, one can see that DFT-PBE, without vdW corrections, lies quite far from the experimental data. Better agreement is achieved with DFT-D contributions, especially DFT-D2. 
The upper branch of the spectrum is particularly well reproduced in this last method. It has to be noticed that the dispersion is quite remarkably similar with all the methods; the frequencies are just
underestimated for example in DFT-PBE by an almost constant factor. 
The better agreement with experimental data of DFT-D2 results, compared with DFT-D3 or DFT-D3(BJ) is primarily due to its better lattice constant.

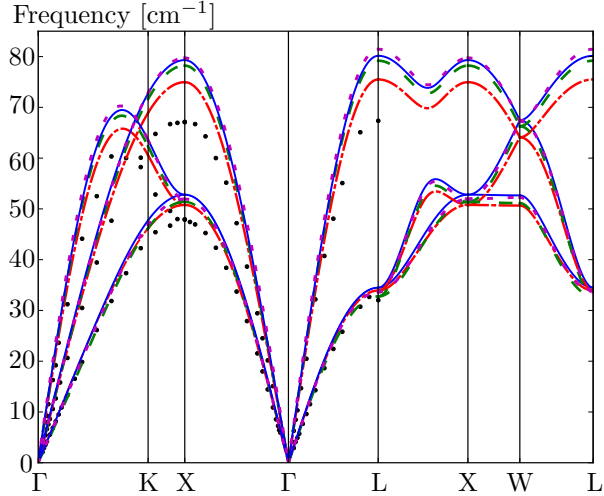
\begin{figure}[h]
\hspace{-0.3cm}\begingroup%
  \makeatletter%
  \providecommand\color[2][]{%
    \errmessage{(Inkscape) Color is used for the text in Inkscape, but the package 'color.sty' is not loaded}%
    \renewcommand\color[2][]{}%
  }%
  \providecommand\transparent[1]{%
    \errmessage{(Inkscape) Transparency is used (non-zero) for the text in Inkscape, but the package 'transparent.sty' is not loaded}%
    \renewcommand\transparent[1]{}%
  }%
  \providecommand\rotatebox[2]{#2}%
  \ifx\svgwidth\undefined%
    \setlength{\unitlength}{250bp}%
    \ifx\svgscale\undefined%
      \relax%
    \else%
      \setlength{\unitlength}{\unitlength * \real{\svgscale}}%
    \fi%
  \else%
    \setlength{\unitlength}{\svgwidth}%
  \fi%
  \global\let\svgwidth\undefined%
  \global\let\svgscale\undefined%
  \makeatother%
  \begin{picture}(1,0.75384615)%
    \put(0,0){\includegraphics[width=\unitlength]{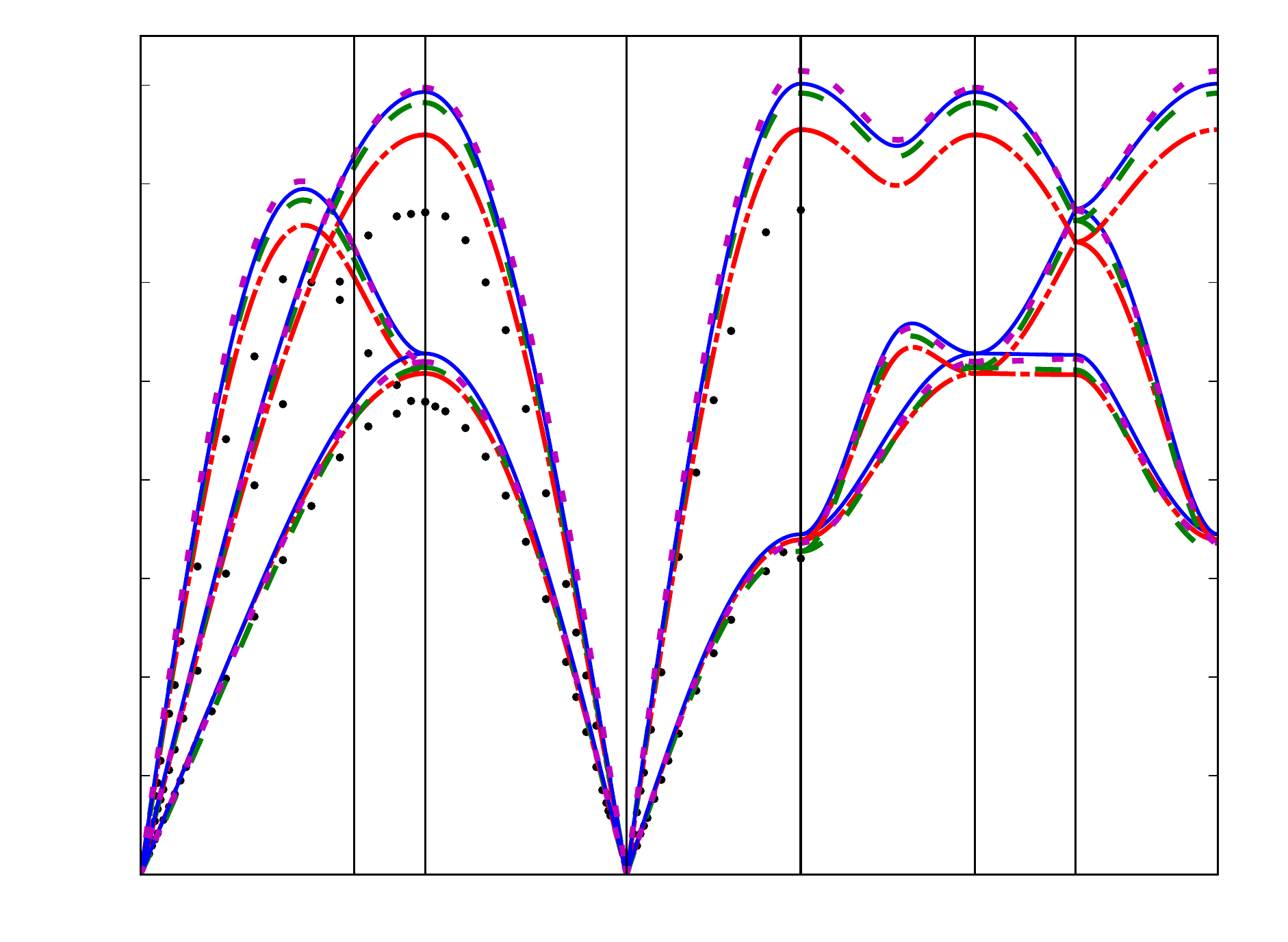}}%
    \put(0.09901709,0.010){\makebox(0,0)[lb]{$\Gamma$}}%
    \put(0.26047863,0.010){\makebox(0,0)[lb]{K}}%
    \put(0.31721368,0.010){\makebox(0,0)[lb]{X}}%
    \put(0.47453504,0.010){\makebox(0,0)[lb]{$\Gamma$}}%
    \put(0.61102222,0.010){\makebox(0,0)[lb]{L}}%
    \put(0.74259658,0.010){\makebox(0,0)[lb]{X}}%
    \put(0.81502222,0.010){\makebox(0,0)[lb]{W}}%
    \put(0.93619316,0.010){\makebox(0,0)[lb]{L}}%
    \put(0.0818051,0.03877402){\makebox(0,0)[lb]{0}}%
    \put(0.0651650103,0.11772427){\makebox(0,0)[lb]{10}}%
    \put(0.06519575316,0.19567521){\makebox(0,0)[lb]{20}}%
    \put(0.06519583333,0.27162393){\makebox(0,0)[lb]{30}}%
    \put(0.06519527248,0.34757436){\makebox(0,0)[lb]{40}}%
    \put(0.06519583333,0.42352479){\makebox(0,0)[lb]{50}}%
    \put(0.06519569983,0.49447521){\makebox(0,0)[lb]{60}}%
    \put(0.06519594017,0.579542564){\makebox(0,0)[lb]{70}}%
    \put(0.06519564632,0.65237607){\makebox(0,0)[lb]{80}}%
    \put(0.07108769,0.705547009){\makebox(0,0)[lb]{Frequency [cm$^{-1}$]}}%
  \end{picture}%
\endgroup%
\caption{\label{PBS_Argon3} Phonon band structure of solid $^{36}$Ar computed with the different DFT-D corrections at the experimental lattice parameter. \textsc{Dotted purple:} DFT-PBE without DFT-D correction; 
\textsc{Dashed green:} DFT-D3; \textsc{Solid blue:} \textsc{DFT-D3(BJ)}, \textsc{Dash-dot red:} \textsc{DFT-D2}. Experimental data from \textsc{Ref.} \onlinecite{Fujii1974} are also shown (black dots). }
\end{figure}

Finally, we computed the phonon band structure at experimental lattice constant 5.30017 \AA $\,$ \cite{Peterson1966} to get further insights on the direct effect on the DFT-D methods on the phonon frequencies. Indeed,
these methods give a quite different lattice parameter for argon. The results are shown in \textsc{Fig. \ref{PBS_Argon3}}.

One can see that in all the cases, the phonon band structures overestimate the experimental results. It seems once again that DFT-D2 improves the agreement with the experiments while DFT-D3 
and DFT-D3(BJ) lies quite close to the DFT-PBE phonon band structure. Further discussions on the importance of the contribution to the IFCs of the DFT-D corrections can be found in \textsc{S.I.}.

\subsection{Graphite}
In graphite, each layer is bound to the neighboring ones by weak forces.
Therefore, this material requires proper description of these interactions in DFT. We focus in this work on AB-stacked graphite.

The interlayer distance predicted 
in PBE (4.4 \AA) largely overestimates the experimental value 
of 3.34 \AA $\,$ \cite{Baskin1955}. The use of DFT-D2 gives an interlayer distance of 3.21 \AA $\,$ and thus tends to overestimate the binding force between the layers. 
DFT-D3 and DFT-D3(BJ) predict 3.48 \AA $\,$ and 3.37 \AA $\,$ distances, respectively, within 4\% and 1\% of the experiments, respectively. The values of the in-plane lattice parameter and cohesive energy can be found
in \textsc{Tab. \ref{TabGraphite}}. 

\begin{table}[h]
\begin{center}
\begin{tabular}{l  c  c   c }
\hline \hline
Correction & a [\AA]& d [\AA] & E$_{coh}$ [meV/C]  \\
\hline
DFT-PBE &  2.46 & 4.4 & -1.3  \\
DFT-D2 & 2.45 & 3.21  & -57.8 \\
DFT-D3 & 2.46 & 3.48 &  -48.9 \\
DFT-D3(BJ) & 2.46 & 3.37 & -53.9   \\
\hline
optB88-vdW$^a$ & 2.47 & 3.36 & -69.5   \\
vdW-DF2$^b$ & 2.47 & 3.52 & \\
rVV10$^b$ & 2.46 & 3.36 &\\
\hline
Exp. & 2.4589$^c$ & 3.3538$^c$ & -52$\pm$5$^d$   \\
\hline \hline
$^a$ \textsc{Ref.} \onlinecite{Hazrati2014} & &   \\
$^b$ \textsc{Ref.} \onlinecite{Sabatini2012}, \textsc{Tab. 6-1} & & \\
$^c$ \textsc{Ref.} \onlinecite{Baskin1955} & &   \\
$^d$ \textsc{Ref.} \onlinecite{Zacharia2004} & &   \\
\end{tabular}
\caption{Lattice parameters and graphite cohesive energy per carbon atom computed with the different DFT-D methods. Results obtained by Hazrati, Sabatini
and their respective co-workers with the family of vdW-DF methods \cite{Hazrati2014,Sabatini2012},
as well as experimental data are also shown.\label{TabGraphite}}
\end{center}
\end{table}

\begin{figure}[h]
\hspace{0.3cm}\begingroup%
  \makeatletter%
  \providecommand\color[2][]{%
    \errmessage{(Inkscape) Color is used for the text in Inkscape, but the package 'color.sty' is not loaded}%
    \renewcommand\color[2][]{}%
  }%
  \providecommand\transparent[1]{%
    \errmessage{(Inkscape) Transparency is used (non-zero) for the text in Inkscape, but the package 'transparent.sty' is not loaded}%
    \renewcommand\transparent[1]{}%
  }%
  \providecommand\rotatebox[2]{#2}%
  \ifx\svgwidth\undefined%
    \setlength{\unitlength}{250.51635742bp}%
    \ifx\svgscale\undefined%
      \relax%
    \else%
      \setlength{\unitlength}{\unitlength * \real{\svgscale}}%
    \fi%
  \else%
    \setlength{\unitlength}{\svgwidth}%
  \fi%
  \global\let\svgwidth\undefined%
  \global\let\svgscale\undefined%
  \makeatother%
  \begin{picture}(1,0.769084)%
    \put(0,0){\includegraphics[width=\unitlength]{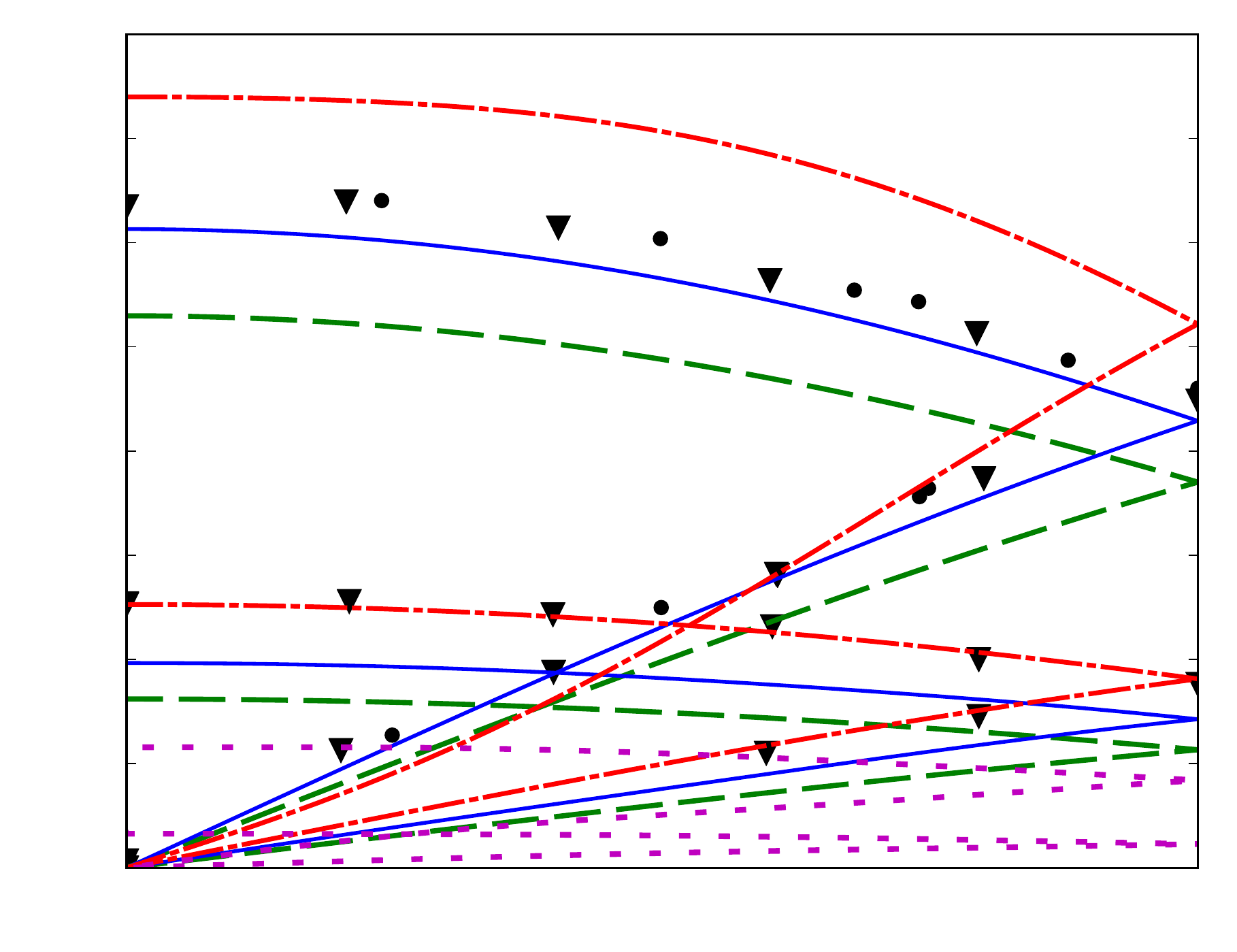}}%
    \put(0.09480157,0.02784536){\makebox(0,0)[lb]{$\Gamma$}}%
    \put(0.95380157,0.02784536){\makebox(0,0)[lb]{A}}%
    \put(0.072232137,0.05585788){\makebox(0,0)[lb]{0}}%
    \put(0.05287579,0.14009593){\makebox(0,0)[lb]{20}}%
    \put(0.05233855,0.22433398){\makebox(0,0)[lb]{40}}%
    \put(0.05281609,0.30857204){\makebox(0,0)[lb]{60}}%
    \put(0.0527564,0.39281009){\makebox(0,0)[lb]{80}}%
    \put(0.03911667,0.47704815){\makebox(0,0)[lb]{100}}%
    \put(0.03911667,0.56128621){\makebox(0,0)[lb]{120}}%
    \put(0.03911667,0.64552426){\makebox(0,0)[lb]{140}}%
    \put(0.03911667,0.72976232){\makebox(0,0)[lb]{160}}%
    \put(0.10826592,0.73800349){\makebox(0,0)[lb]{Frequency [cm$^{-1}$]}}%
  \end{picture}%
\endgroup%
\caption{\label{PBS_Graphite} Phonon band structure of low-frequency modes of graphite computed with DFT-PBE as well as adding to it different DFT-D corrections; \textsc{Dotted purple:} computed without DFT-D corrections; 
\textsc{Dash-dot red:} computed with DFT-D2; \textsc{Dashed green:} computed with DFT-D3; \textsc{Solid blue:} computed with DFT-D3(BJ).
In each case, the computations were performed at the corresponding relaxed lattice parameters. 
Experimental data from \textsc{Ref.} \onlinecite{Mohr2007}  (black dots) and \textsc{Ref.} \onlinecite{Nicklow1972} (black triangles) are also shown. }
\end{figure}
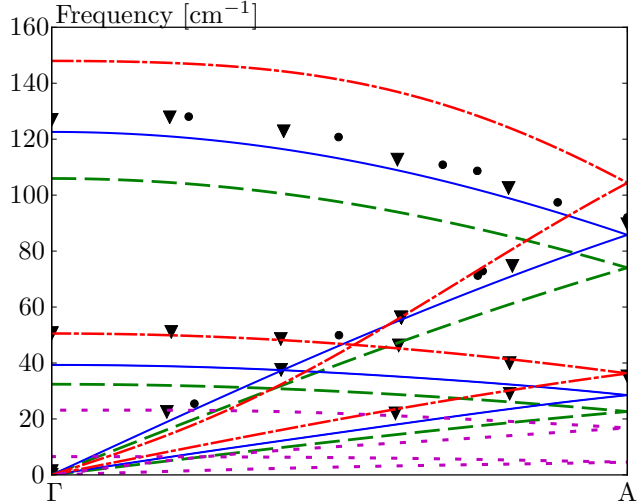

The graphite lattice phonon bands along $\Gamma-A$, with the different methods, are shown on \textsc{Fig. \ref{PBS_Graphite}}. 
For an easier comparison, the lattice phonon frequencies at $\Gamma$ and $A$ are reported in \textsc{Tab. \ref{TabGraphite2}} alongside
with experimental data (at room temperature) and results obtained with vdW-DF2 (and other optimized functionals for the vdW). 

\begin{table}[h]
\begin{center}
\begin{tabular}{l  c  c   c  c}
\hline \hline
Frequencies & $\Gamma_{LO}$ & $\Gamma_{ZO}$ & $A_{TA/TO}$ & $A_{LA/LO}$  \\
\hline
DFT-PBE& 6.5 & 23 & 4.5 & 17 \\
DFT-D2 & 51 & 148 & 36 & 104  \\
DFT-D3 & 32 & 106 & 23 &74  \\
DFT-D3(BJ) & 39 & 123 & 29 & 86  \\
\hline
optB88-vdW$^a$ & 40 & 139 & 28 & 95  \\
rVV10$^b$ & 41 & 140 & 29 & 98  \\
vdW-DF2$^b$ & 31 & 118 & 22 & 82  \\
\hline
Exp.$^c$ & 49 & 126 & 35 & 89  \\
Exp.$^d$ & 42 & 127 & & \\
\hline \hline
$^a$ \textsc{Ref.} \onlinecite{Hazrati2014} & & & & \\
$^b$ \textsc{Ref.} \onlinecite{Sabatini2012}, \textsc{Fig. 6-3} & & & & \\
$^c$ \textsc{Ref.} \onlinecite{Mohr2007}, \textsc{Fig. 4} & & & & \\
$^d$ \textsc{Ref.} \onlinecite{Mohr2007}, p.2 & & & & \\
\end{tabular}
\caption{Low phonon frequencies of graphite computed with different methods to treat the dispersion and at special points of the reciprocal space.
Results obtained by Hazrati and co-workers with optB88-vdW \cite{Hazrati2014}, by Sabatini \cite{Sabatini2012} are also presented as well as experimental data from Nicklow and co-workers \cite{Nicklow1972}. \label{TabGraphite2}}
\end{center}
\end{table}

As one can see, the lattice modes are underestimated in DFT-PBE (largely) and in DFT-D3 with respect to
experiments. DFT-D2 describes much better the $\Gamma_{LO}$ but overestimates by more than 20 cm$^{-1}$ the $\Gamma_{ZO}$ mode. DFT-D3(BJ) works better to describe
the higher lattice branch -although being as poor as the other methods for the $\Gamma_{LO}$ one- and yields similar precision than the more sophisticated vdW-DF2 methods (and other optimized functionals for the vdW) to describe these lattice modes. 
These discrepancies may be explained by the choice of the exchange-correlation approximation but perhaps as well by temperature effects (experimental data are performed at room temperature).

\begin{figure}[h]
\hspace{0.3cm}\begingroup%
  \makeatletter%
  \providecommand\color[2][]{%
    \errmessage{(Inkscape) Color is used for the text in Inkscape, but the package 'color.sty' is not loaded}%
    \renewcommand\color[2][]{}%
  }%
  \providecommand\transparent[1]{%
    \errmessage{(Inkscape) Transparency is used (non-zero) for the text in Inkscape, but the package 'transparent.sty' is not loaded}%
    \renewcommand\transparent[1]{}%
  }%
  \providecommand\rotatebox[2]{#2}%
  \ifx\svgwidth\undefined%
    \setlength{\unitlength}{250.51635742bp}%
    \ifx\svgscale\undefined%
      \relax%
    \else%
      \setlength{\unitlength}{\unitlength * \real{\svgscale}}%
    \fi%
  \else%
    \setlength{\unitlength}{\svgwidth}%
  \fi%
  \global\let\svgwidth\undefined%
  \global\let\svgscale\undefined%
  \makeatother%
  \begin{picture}(1,0.769084)%
    \put(0,0){\includegraphics[width=\unitlength]{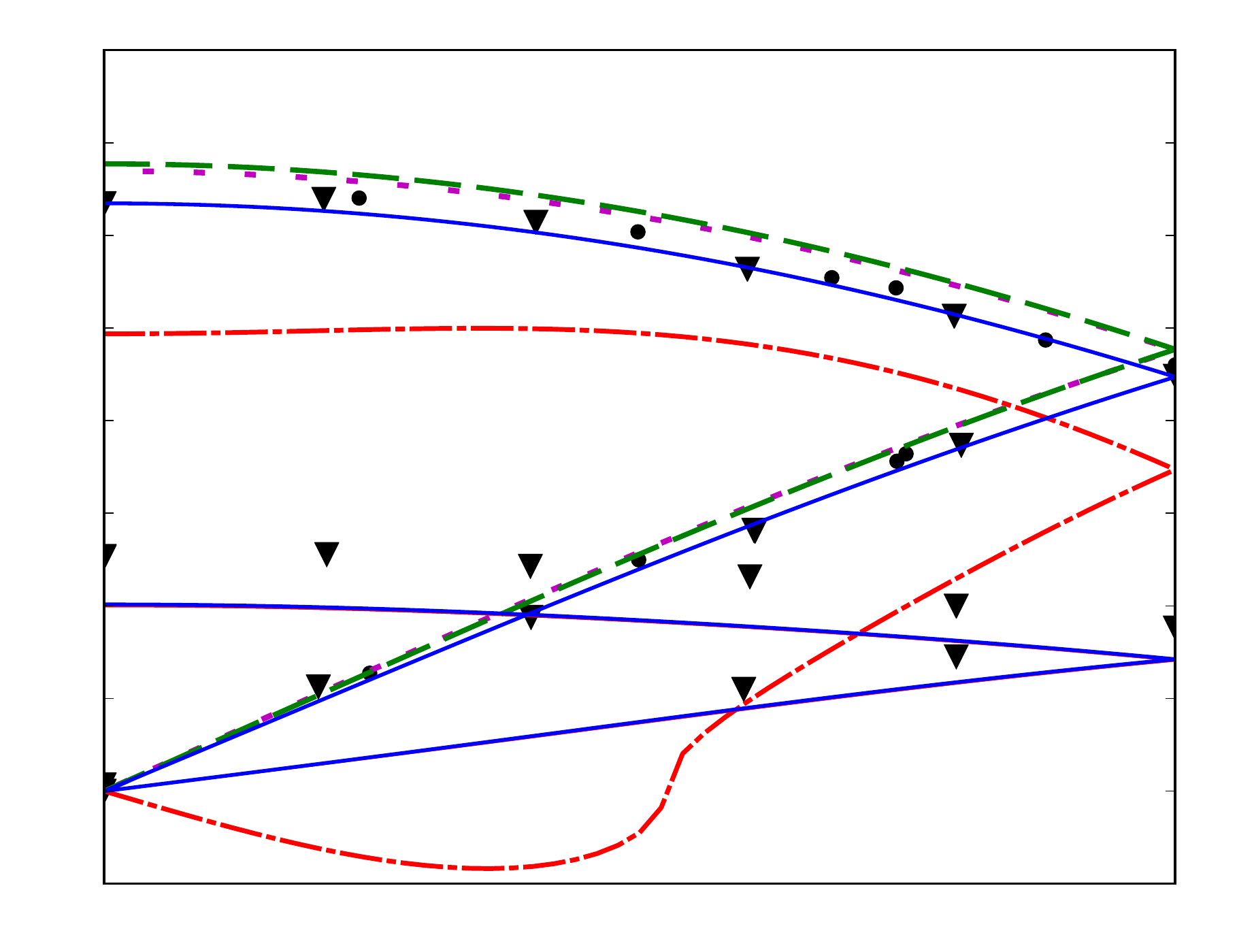}}%
    \put(0.0727564,0.02084536){\makebox(0,0)[lb]{$\Gamma$}}%
    \put(0.93427564,0.02084536){\makebox(0,0)[lb]{A}}%
    \put(0.05727564,0.119585788){\makebox(0,0)[lb]{0}}%
    \put(0.0427564,0.195009593){\makebox(0,0)[lb]{20}}%
    \put(0.0427564,0.26733398){\makebox(0,0)[lb]{40}}%
    \put(0.0427564,0.342857204){\makebox(0,0)[lb]{60}}%
    \put(0.0427564,0.41881009){\makebox(0,0)[lb]{80}}%
    \put(0.02411667,0.49704815){\makebox(0,0)[lb]{100}}%
    \put(0.02411667,0.57128621){\makebox(0,0)[lb]{120}}%
    \put(0.02411667,0.64552426){\makebox(0,0)[lb]{140}}%
    \put(0.02411667,0.71676232){\makebox(0,0)[lb]{160}}%
    \put(0.09326592,0.73100349){\makebox(0,0)[lb]{Frequency [cm$^{-1}$]}}%
  \end{picture}%
\endgroup%
\caption{\label{PBS_Graphite2} Phonon band structure of low-frequency modes of graphite computed for different methods on top of PBE at experimental lattice parameters \cite{Baskin1955}; \textsc{Dotted purple:} computed without DFT-D corrections; 
\textsc{Dash-dot red:} computed with DFT-D2; \textsc{Green:} computed with DFT-D3; \textsc{Solid blue:} computed with DFT-D3(BJ).
Experimental data from \textsc{Ref.} \onlinecite{Mohr2007}  (black dots) and \textsc{Ref.} \onlinecite{Nicklow1972} (black triangles) are also shown. }
\end{figure}

In addition, we computed at experimental lattice constants \cite{Baskin1955} the phonon branches along $\Gamma-A$ with the different DFT-D methods. The results are shown in \textsc{Fig. \ref{PBS_Graphite2}}.

DFT-PBE and DFT-D3 give quite similar results while DFT-D3(BJ) lies closer to the experiments for the high frequency lattice branch. DFT-D2 is completely off, predicting negative phonon frequencies,
which typically indicates a phase instability. These negative phonon modes have likely the same origin than the underestimation in DFT-D2 of the outplane lattice constant compared to the experiments.

\subsection{Benzene}
Finally, we studied the benzene molecular crystal. This material crystallizes at 5.5$^\circ$C; its primitive cell is orthorhombic (Pbca space group) and contains four C$_6$H$_6$ molecules. 

With DFT-PBE, we observed a large overestimation
of the experimental volume \cite{Kohzin1954} by more than 30\%.
This can be explained by the fact that the benzene molecules are bound by vdW interactions in the crystal, which are not included -or somehow spuriously- in PBE. So, it was not meaningful to compute the phonon frequencies in this case.
In contrast, relaxations performed with the different DFT-D methods yield a meaningful global energy minimum; the lattice parameters obtained with each method are given in \textsc{Tab. \ref{TabABenz}},
as well as their corresponding cohesive energy for the benzene crystal. In this table, we report as well the low-temperature experimental data (77 K) $a=7.292$ \AA, $b =9.471$ \AA $\,$
and $c=6.742$ \AA $\,$ \cite{Kohzin1954} and experimental measurements performed by Jeffrey and co-workers at 10K on deuterated benzene \cite{Jeffrey1987} $a=7.360$ \AA, 
$b=9.375$ \AA $\,$ and $c=6.703$ \AA. Concerning the cohesive energy, an estimate of the lattice energy at 0 K \cite{Yang2014} of -55.3$\pm$ 2.2 kJ/mol is also shown.

\begin{table}[h]
\begin{center}
\begin{tabular}{l  c  c   c c}
\hline \hline
Correction & a [\AA]& b [\AA] & c [\AA] & E$_{coh}$ [kJ/mol]  \\
\hline
DFT-D2 & 7.13 & 9.07  & 6.44 & -56.3\\
DFT-D3 & 7.42 & 9.48 &  6.85  & -47.4 \\
DFT-D3(BJ) & 7.30 & 9.31 & 6.70 & -55.0   \\
\hline
Exp.$^a$ & 7.292  & 9.471 & 6.742   \\
Exp.$^b$ & 7.360  & 9.375 & 6.703   \\
Est.$^c$ & & & & -55.3$\pm$ 2.2 \\
\hline \hline
$^a$ \textsc{Ref.} \onlinecite{Kohzin1954} & &   \\
$^b$ \textsc{Ref.} \onlinecite{Jeffrey1987} & &   \\
$^c$ \textsc{Ref.} \onlinecite{Yang2014} & &   \\
\end{tabular}
\caption{Benzene lattice parameters and cohesive energy computed with the different DFT-D methods. Available experimental data for the lattice parameters and  an estimated value of
the lattice energy at 0K are also reported.\label{TabABenz}}
\end{center}
\end{table}

As one can see, DFT-D3 and DFT-D3(BJ) improve markedly the agreement with the experiments, with the later method providing the better description of benzene molecular crystal.
Our DFT-D3(BJ) implementation predicts a cohesive energy of -55.0 kJ/mol in agreement both with the theoretical value of -55.0 kJ/mol reported with the same method \cite{Moellemann2014} 
and with the estimate of the lattice energy at 0 K.
In the case of DFT-D2, the results are more contrasted, with the $b$ and $c$ lattice parameters being quite underestimated in this method. These discrepancies probably originate from the absence of any dependence
of the dispersion coefficient on the close chemical environment around each atom in this method.

 As there are three translational and rotational degrees of freedom for each molecule,
there are in total 24 lattice modes, including the three acoustic modes.
The computed lattice frequencies of benzene at the zone-center are reported in \textsc{Tab. \ref{TabLattBenzene}} alongside with experimental Raman measurements performed at 7 K \cite{Pinan1998}. 

As one can see, DFT-D3(BJ) is able to reproduce quite well the experimental frequencies of the lattice modes. On average, the difference is less than 5 cm$^{-1}$ compared to the experiments 
which is quite acceptable from this degree of theory.
Only the $B_{1g}$ mode with 107.3 cm$^{-1}$ experimental frequency, which corresponds to a rotation of the phenyl groups in opposite phases, is relatively poorly described.
The discrepancies may arise from several sources: they may come from the choice of the functional approximation (including the vdW method), from the choice of pseudopotentials or from anharmonicity effects.
We also observed only weak LO-TO splitting for the lattice modes; the most important effect in DFT-D3(BJ) is seen for the B$_{1u}$ at 103 cm$^{-1}$ that is shifted by 0.6 cm$^{-1}$ upwards 
for a non-analyticity alongside c$^*$ axis.

\begin{table}[h]
\begin{center}
\begin{tabular}{l c c c c  | l c c c}
\hline \hline
Sym. &  \multicolumn{4}{c|}{Frequencies} & Sym. & \multicolumn{3}{c}{Frequencies} \\
\cline{2-5} \cline{7-9}

 & -D2 & -D3 &-BJ& Exp.$^a$&  &-D2& -D3&-BJ  \\
\hline
$A_g$ &78 & 54 &60 & 63.3 & $B_{2u}$&75& 52 &59 \\
$B_{1g}$ &81& 60& 69 & 67.0 & $A_u$&68& 52 &60 \\
$B_{3g}$ &86& 55 &64& 68.7 & $B_{1u}$ &88&62&71\\
$B_{2g}$ &99&75 &88 & 84.9 & $A_{u}$ &89&62&72\\
$A_g$&99& 74 &83 & 85.0 &  $B_{3u}$ &89&67&76 \\
$B_{2g}$ &118& 82 &93 & 89.4 &$B_{3u}$&110&80&91 \\
$B_{2g}$ &132& 94& 109 & 97.3& $A_u$&121&89& 103 \\
$A_g$ &118& 84 &97 & 100.6& $B_{1u}$&124&90& 104 \\
$B_{3g}$  &120& 88 &103 & 100.6&$B_{2u}$&125&93&107 \\
$B_{1g}$  &107& 77 &88 & 107.3&&  \\
$B_{1g}$ &160& 120 &136& 135.0&\\
$B_{3g}$ &157& 119 &136& 136.0&& \\
\hline
 MAE [cm$^{-1}$] & 18.5& 12.7 & 4.54 & & &\\
 MAPE [\%]& 20.2 & 13.4 & 4.9 & & \\
\hline \hline
\multicolumn{4}{l}{$^a$ \textsc{Ref.} \onlinecite{Pinan1998}} \\
\end{tabular}

\caption{Lattice phonon frequencies of benzene molecular crystal computed at the zone-center wavevector with PBE-D2, PBE-D3 and PBE-D3(BJ). Experimental data from Pinan and co-workers \cite{Pinan1998} are also presented for Raman active modes,
as well as mean average error (MAE) and mean average pourcentage error (MAPE) for each given method with respect to the experiments. \label{TabLattBenzene}}
\end{center}
\end{table}

%
%
%

\begin{figure}[h]
\begin{center}
\hspace{-0.3cm}\begingroup%
  \makeatletter%
  \providecommand\color[2][]{%
    \errmessage{(Inkscape) Color is used for the text in Inkscape, but the package 'color.sty' is not loaded}%
    \renewcommand\color[2][]{}%
  }%
  \providecommand\transparent[1]{%
    \errmessage{(Inkscape) Transparency is used (non-zero) for the text in Inkscape, but the package 'transparent.sty' is not loaded}%
    \renewcommand\transparent[1]{}%
  }%
  \providecommand\rotatebox[2]{#2}%
  \ifx\svgwidth\undefined%
    \setlength{\unitlength}{250bp}%
    \ifx\svgscale\undefined%
      \relax%
    \else%
      \setlength{\unitlength}{\unitlength * \real{\svgscale}}%
    \fi%
  \else%
    \setlength{\unitlength}{\svgwidth}%
  \fi%
  \global\let\svgwidth\undefined%
  \global\let\svgscale\undefined%
  \makeatother%
  \begin{picture}(1,0.75384615)%
    \put(0,0){\includegraphics[width=\unitlength]{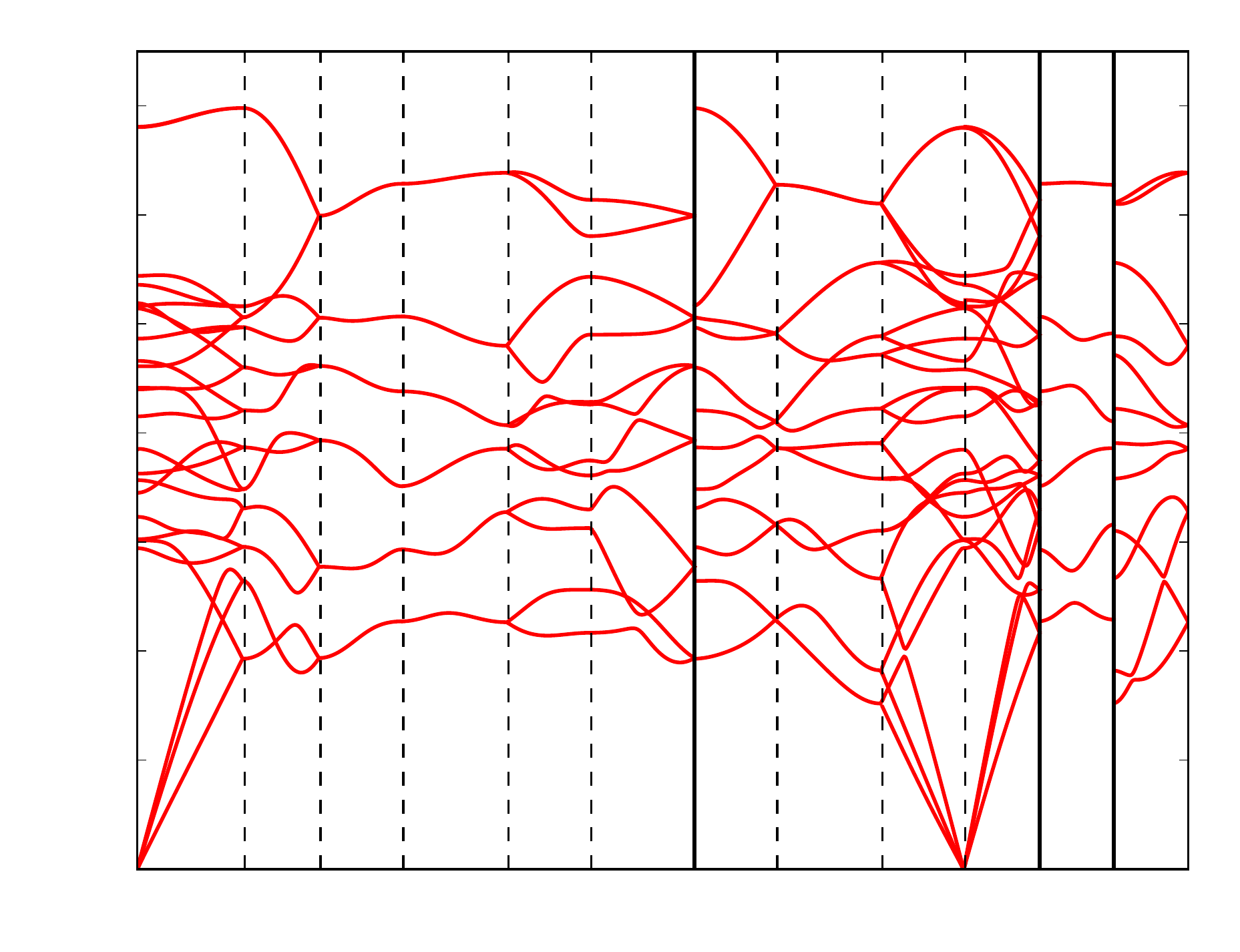}}%
    \put(0.1076239,0.03452457){\makebox(0,0)[lb]{$\Gamma$}}%
    \put(0.18585205,0.03358974){\makebox(0,0)[lb]{X}}%
    \put(0.24547626,0.03358974){\makebox(0,0)[lb]{U}}%
    \put(0.31330704,0.03358974){\makebox(0,0)[lb]{R}}%
    \put(0.39752132,0.03358974){\makebox(0,0)[lb]{T}}%
    \put(0.46668598,0.03358974){\makebox(0,0)[lb]{Z}}%
    \put(0.52865672,0.02558974){\makebox(0,0)[lb]{U$|$X}}%
    \put(0.61761945,0.03358974){\makebox(0,0)[lb]{S}}%
    \put(0.69825949,0.03358974){\makebox(0,0)[lb]{Y}}%
    \put(0.76768962,0.03358974){\makebox(0,0)[lb]{$\Gamma$}}%
    \put(0.81049786,0.02558974){\makebox(0,0)[lb]{Z$|$R}}%
    \put(0.8718881624,0.02558974){\makebox(0,0)[lb]{S$|$Y}}%
    \put(0.94368323,0.03358974){\makebox(0,0)[lb]{T}}%
    \put(0.0882585,0.05891934){\makebox(0,0)[lb]{0}}%
    \put(0.0678985,0.14532959){\makebox(0,0)[lb]{20}}%
    \put(0.0678985,0.23173985){\makebox(0,0)[lb]{40}}%
    \put(0.0678985,0.3191501){\makebox(0,0)[lb]{60}}%
    \put(0.0678985,0.405956036){\makebox(0,0)[lb]{80}}%
    \put(0.0488985,0.495997062){\makebox(0,0)[lb]{100}}%
    \put(0.0488985,0.58238087){\makebox(0,0)[lb]{120}}%
    \put(0.0488985,0.67279113){\makebox(0,0)[lb]{140}}%
    \put(0.10840064,0.73061538){\makebox(0,0)[lb]{Frequency [cm$^{-1}$]}}
  \end{picture}%
\endgroup%
\caption{\label{Phonon_Structure} Lattice phonon band structure of benzene molecular crystal (Pbca space group) computed with PBE-D3(BJ).}
\end{center}
\end{figure}
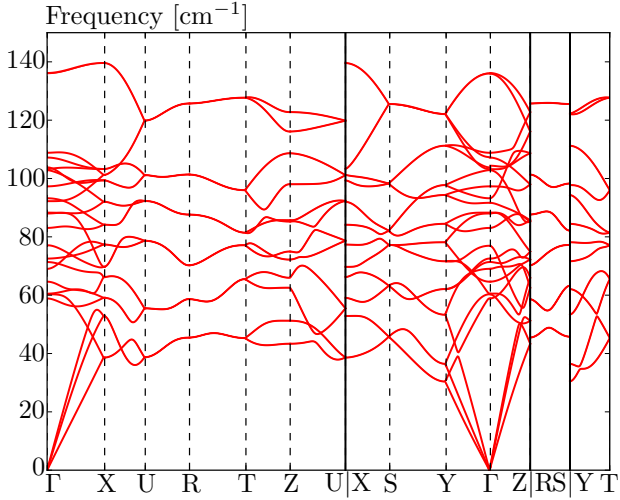

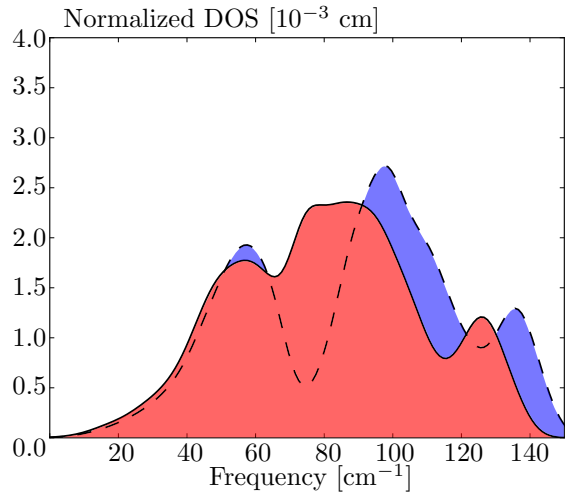
\begin{figure}[h]
\begin{center}
\hspace{-0.3cm}\begingroup%
  \makeatletter%
  \providecommand\color[2][]{%
    \errmessage{(Inkscape) Color is used for the text in Inkscape, but the package 'color.sty' is not loaded}%
    \renewcommand\color[2][]{}%
  }%
  \providecommand\transparent[1]{%
    \errmessage{(Inkscape) Transparency is used (non-zero) for the text in Inkscape, but the package 'transparent.sty' is not loaded}%
    \renewcommand\transparent[1]{}%
  }%
  \providecommand\rotatebox[2]{#2}%
  \ifx\svgwidth\undefined%
    \setlength{\unitlength}{250.19667969bp}%
    \ifx\svgscale\undefined%
      \relax%
    \else%
      \setlength{\unitlength}{\unitlength * \real{\svgscale}}%
    \fi%
  \else%
    \setlength{\unitlength}{\svgwidth}%
  \fi%
  \global\let\svgwidth\undefined%
  \global\let\svgscale\undefined%
  \makeatother%
  \begin{picture}(1,0.78941033)%
    \put(0,0){\includegraphics[width=\unitlength]{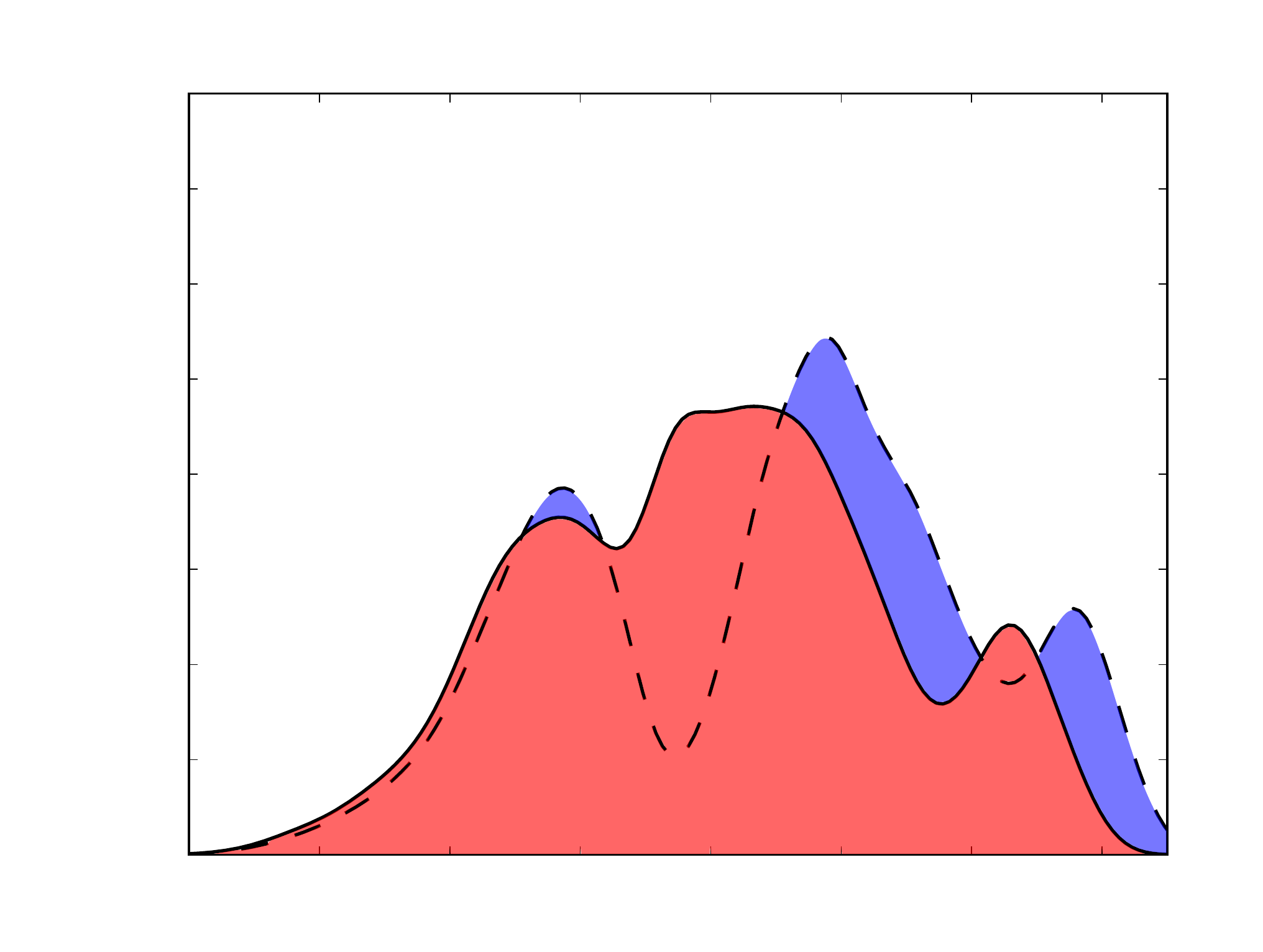}}%
    \put(0.09371699,0.66626975){\color[rgb]{0,0,0}\makebox(0,0)[lb]{\smash{4.0}}}%
    \put(0.09371699,0.59465171){\color[rgb]{0,0,0}\makebox(0,0)[lb]{\smash{3.5}}}%
    \put(0.09371699,0.51826452){\color[rgb]{0,0,0}\makebox(0,0)[lb]{\smash{3.0}}}%
    \put(0.09371699,0.44383294){\color[rgb]{0,0,0}\makebox(0,0)[lb]{\smash{2.5}}}%
    \put(0.09371699,0.365102064){\color[rgb]{0,0,0}\makebox(0,0)[lb]{\smash{2.0}}}%
    \put(0.09371699,0.29176236){\color[rgb]{0,0,0}\makebox(0,0)[lb]{\smash{1.5}}}%
    \put(0.09371699,0.21439612){\color[rgb]{0,0,0}\makebox(0,0)[lb]{\smash{1.0}}}%
    \put(0.09371699,0.14243518){\color[rgb]{0,0,0}\makebox(0,0)[lb]{\smash{0.5}}}%
    \put(0.09371699,0.04892023){\color[rgb]{0,0,0}\makebox(0,0)[lb]{\smash{0.0}}}%
    \put(0.23761293,0.04176654){\color[rgb]{0,0,0}\makebox(0,0)[lb]{\smash{20}}}%
    \put(0.34129555,0.04176654){\color[rgb]{0,0,0}\makebox(0,0)[lb]{\smash{40}}}%
    \put(0.44358197,0.04176654){\color[rgb]{0,0,0}\makebox(0,0)[lb]{\smash{60}}}%
    \put(0.54947847,0.04176654){\color[rgb]{0,0,0}\makebox(0,0)[lb]{\smash{80}}}%
    \put(0.63954468,0.04176654){\color[rgb]{0,0,0}\makebox(0,0)[lb]{\smash{100}}}%
    \put(0.7432378,0.04176654){\color[rgb]{0,0,0}\makebox(0,0)[lb]{\smash{120}}}%
    \put(0.84252419,0.04176654){\color[rgb]{0,0,0}\makebox(0,0)[lb]{\smash{140}}}%
    \put(0.38977138,0.0052564373){\color[rgb]{0,0,0}\makebox(0,0)[lb]{\smash{Frequency [cm$^{-1}$]}}}%
    \put(0.17050542,0.695119132){\color[rgb]{0,0,0}\makebox(0,0)[lb]{\smash{Normalized DOS [10$^{-3}$ cm]}}}%
  \end{picture}%
\endgroup%
\caption{\label{Exp_PDOS_Benz} Comparison between lattice phonon density of states of benzene molecular crystal computed with DFT-D3(BJ) at
the experimental lattice parameters \cite{Jeffrey1987} (blue-filled dotted line) or at the DFT-D3(BJ) relaxed ones (plain curve filled with red). }
\end{center}
\end{figure}

In addition, we computed the phonon band structure and phonon density of states of this material with DFT-D3(BJ). 
The phonon band structure and the phonon DOS for the lattice modes are presented in \textsc{Fig. \ref{Phonon_Structure}} and \textsc{Fig. \ref{Exp_PDOS_Benz}}, respectively. 
As one can see in \textsc{Fig. \ref{Phonon_Structure}}, the 24 lattice modes merge into 6 modes along U-R, R-T and R-S branches.
We observe that the phonon density of states computed at relaxed DFT-D3(BJ) has the following peak maxima: 57, 78, 87 and 126 cm$^{-1}$ which are in reasonable
agreement with the experimental peaks reported by Pinan and coworkers in the double-resonance Raman spectrum \cite{Pinan1998} i.e. at around 40, 80, 90, and 120 cm$^{-1}$. 
 It has to be noticed, though, that the discrepancies may arise from anharmonic effects, present in the experiments but that we neglect 
in our computation. An additional graph presented in \textsc{S.I.}, \textsc{Fig. S2}, compares the phonon density of states computed at DFT-D3(BJ), with and without the DFT-D3(BJ) correction.

Finally, the lattice phonon density of states computed at experimental lattice parameters and internal positions \cite{Jeffrey1987} with DFT-D3(BJ) is also shown in \textsc{Fig. \ref{Exp_PDOS_Benz}}.
As one can see, working at experimental parameters does not
improve globally the agreement with the experiments, as only three peaks are observed in that case, in contradiction with the experiments. 

\section*{Conclusion}

We have presented in this work the theoretical derivation of the pair-wise part of the DFT-D contribution to the IFCs and dynamical matrices, as well as
its implementation inside the \textsc{Abinit} software. We have validated the implementation with respect to frozen-phonon computations, and also tested the hypothesis that the contribution from three-body interactions can be neglected. We have then applied this new implementation to the computation of the phonon band structures
of argon, graphite and benzene materials, that are known to require proper description of the long-range e$^-$-e$^-$ correlation. We have analyzed the specific role of the correctness of the equilibrium parameters, and the one of the
direct modification of dynamical matrices by the DFT-D contribution.

For argon, all the DFT-D methods improve markedly over the DFT-PBE results. An excellent agreement with experimental data is even obtained for the DFT-D2 method, taken at its relaxed lattice parameter. The agreement for the DFT-D3 and DFT-D3(BJ), 
again at their relaxed lattice parameter is less satisfactory, but still within 10-20\% of the experiment. If one works at the experimental equilibrium lattice parameter, all DFT-D (or DFT-PBE) methods overestimate the phonon frequencies,  by a few percent, the best agreement being again obtained with DFT-D2.

For graphite, at the corresponding relaxed lattice parameters, all DFT-D methods also improve enormously  with respect to DFT-PBE for the description of the low-lying bands, the DFT-D3(BJ) or the DFT-D2 being the best depending on the considered branch. The agreement is again still within 10-20\% of the experiment for these low-lying bands, for all DFT-D methods. When fixing the lattice parameter at the experimental value,  DFT-PBE, DFT-D3 and DFT-D3(BJ) methods give very similar results, while DFT-D2 is considerably off, and even predict instabilities of the lattice. 

For benzene, for the 12 low-lying modes at $\Gamma$ for which experimental data (Raman) is available, spanning the range  between 63.3 cm$^{-1}$ and 163 cm$^{-1}$, the DFT-D3(BJ) has a maximum discrepancy of 19.2 cm$^{-1}$, while ten modes are obtained within 5 cm$^{-1}$ of the experimental values. The performance of the DFT-D3 method are less satisfactory, but still reasonable.

Globally, these dispersive contributions to the IFCs can not neglected and are important to properly reproduce experimental results. Overall the DFT-D3(BJ) is the most reliable from what we observed.

This work opens the way for the computation of more advanced response properties of molecular crystals in DFPT,
like Raman spectra, temperature dependence of electronic properties within the Allen-Heine-Cardona formalism \cite{Allen1983,Gonze2011,Ponce2014,Ponce2014b,Ponce2015,Antonius2014,Antonius2015}
or thermodynamic properties within the quasiharmonic approximation \cite{Born1988,Rignanese1996,Lichtenstein2000,Lazzeri2002,Carrier2008}. 
Further developments may include
the derivation of strain perturbation for the DFT-D methods, as well as the developments of dispersive contributions 
beyond pairwise-additive models \cite{Reilly2015} which should describe in a more adequate way the many-body nature of the vdW interactions.

\section*{Aknowledgments}

The authors acknowledge technical help from J.-M. Beuken and M. Giantomassi. This work was supported by the FRS-FNRS through a FRIA
Grant (B.V.T.) and by the Communauté française de Belgique through the BATTAB project (ARC 14/19-057). Computational resources have been provided by the supercomputing facilities of
the Université catholique de Louvain (CISM/UCL) and the Consortium desThe authors acknowledge technical help from J.-M. Beuken and M. Giantomassi. This work was supported by the FRS-FNRS through a FRIA
Grant (B.V.T.) and by the Communauté française de Belgique through the BATTAB project (ARC 14/19-057). Computational resources have been provided by the supercomputing facilities of
the Université catholique de Louvain (CISM/UCL) and the Consortium des
Equipements de Calcul Intensif en Fédération Wallonie Bruxelles (CECI) funded by
the Fonds de la Recherche Scientifique de Belgique (FRS-FNRS) under convention 2.5020.11.
The present research benefited from computational resources made available on the Tier-1 supercomputer of the Fédération Wallonie-Bruxelles, 
infrastructure funded by the Walloon Region under the grant agreement n$^\circ$1117545.
Equipements de Calcul Intensif en Fédération Wallonie Bruxelles (CECI) funded by
the Fonds de la Recherche Scientifique de Belgique (FRS-FNRS) under convention 2.5020.11.
The present research benefited from computational resources made available on the Tier-1 supercomputer of the Fédération Wallonie-Bruxelles, 
infrastructure funded by the Walloon Region under the grant agreement n$^\circ$1117545.

\end{document}